\documentclass[12pt,letterpaper]{amsart}
\usepackage{graphicx}
\usepackage{subfigure}

\newcommand\twofig{1.6in}
\newcommand\G{\mathcal{G}}
\newcommand\V{\mathcal{V}}
\newcommand\E{\mathcal{E}}
\renewcommand\P{\mathcal{P}} 
\newcommand\abs[1]{\lvert #1 \rvert}
\newcommand\R{\mathbf{R}}
\newcommand\Vss{V_{ss}}
\newcommand\Vrs{V_{rs}}
\newcommand\Vsr{V_{sr}}
\newcommand\Vrr{V_{rr}}
\newcommand\Gs{G_s}
\newcommand\Gr{G_r}

\newcommand\ahat{\hat{a}}
\newcommand\muhat{\hat{\mu}}
\DeclareMathOperator\bias{Bias}
\DeclareMathOperator\mspe{MSPE}
\DeclareMathOperator\rank{Rank}
\DeclareMathOperator\nul{Null}

\DeclareMathOperator\row{Row}

\begin{document}

\title[A Statistical Framework for Efficient Monitoring]{A Statistical Framework for Efficient Monitoring of End-to-End Network
  Properties}
\author[D.\ B.\ Chua]{David B.\ Chua}
\thanks{David B.\ Chua and Eric D.\ Kolaczyk are with the Department of
  Mathematics 
  and Statistics at Boston University. Mark Crovella is with the Department of
  Computer Science at Boston University. 
  Part of this work was performed while E.\ Kolaczyk was with the LIAFA group
  at l'Universit\'e de Paris 7, with support from the Centre National de la
  Recherche Scientifique (CNRS) France, and while M.\ Crovella was at the
  Laboratoire d'Informatique de Paris 6 (LIP6), with support from CNRS and
  Sprint Labs.  This work was supported in part by a grant from Sprint Labs,
  NSF grants ANI-9986397 and CCR-0325701, and by ONR award N000140310043.} 
\author[E.\ D.\ Kolaczyk]{Eric D.\ Kolaczyk}
\author[M.\ Crovella]{Mark Crovella}

\keywords{Algorithms, Statistical analysis, Networking.}

\begin{abstract}
Network service providers and customers are often concerned with aggregate
performance measures that span multiple network paths.  Unfortunately, forming
such network-wide measures can be difficult, due to the issues of scale
involved. In particular, the number of paths grows too rapidly with the number
of endpoints to make exhaustive measurement practical.  As a result, it is of
interest to explore the feasibility of methods that dramatically reduce the
number of paths measured in such situations while maintaining acceptable
accuracy.

In previous work we have proposed a statistical framework for efficiently
addressing this problem, in the context of additive metrics such as delay and
loss rate, for which the per-path metric is a sum of per-link measures (possibly
under appropriate transformation).  The key to our method lies in the
observation and exploitation of the fact that network paths show significant
redundancy (sharing of common links).

In this paper we make three contributions: (1) we generalize the framework to
make it more immediately applicable to network measurements encountered in
practice; (2) we demonstrate that the observed path redundancy upon which our
method is based is robust to variation in key network conditions and
characteristics, including the presence of link failures; and (3) we show how
the framework may be applied to address three practical problems of interest to
network providers and customers, using data from an operating network.  In
particular, we show how appropriate selection of small sets of path measurements
can be used to accurately estimate network-wide averages of path delays, to
reliably detect network anomalies, and to effectively make a choice between
alternative sub-networks, as a customer choosing between two providers or two
ingress points into a provider network.
\end{abstract}

\maketitle

\section{Introduction}
\label{sec:introduction}

In many situations it is important to obtain a network-wide view of path
metrics such as latency and packet loss rate.   For example, in overlay
networks regular measurement of path properties is used to select
alternate routes.  At the IP level, path property measurements
can be used to monitor network health, assess user experience, and
choose between alternate providers, among other applications.
Typical examples of systems performing such measurements include
the NLANR AMP project, the RIPE
Test-Traffic Project, and the Internet
End-to-end Performance Monitoring project \cite{NLANR,RIPETT,iepm}.

Unfortunately extending such efforts to large networks can be difficult,
because the number of network paths grows as the square of the 
number of network endpoints.  Initial work in this area has found that
it is possible to reduce the number of end-to-end measurements to the
number of ``virtual links'' (identifiable link subsets) --- which
typically grows more slowly than the the number of paths ---
and yet still recover the complete set of end-to-end path properties
exactly \cite{chen03:overlay,shavitt01:computing}.

This result is based on a linear algebraic analysis of
routing matrices, as pioneered in \cite{shavitt01:computing}.  A routing
matrix is a binary matrix that specifies 
which links appear in which end-to-end paths.  Such a matrix $G$ has
size (\# paths) $\times$ (\# links), and $G_{i,j} = 1$
if and only if link $j$ is found
along the route taken by path $i$.  The rank of $G$,
which is generally equal to the number of independent paths in the
network,
tends to be much smaller than the total number of paths.  Since a
maximal set of such independent paths can be used to reconstruct any
other path
in the network, it is sufficient to monitor only this set.  
Algorithms for choosing such a set have been 
developed~\cite{shavitt01:computing,chen03:overlay}, 
based on linear algebraic methods of subset selection.

In previous work we have proposed a general approach to extending this
strategy.  First, we note that routing matrices encountered in practice
generally show significant sharing of links between paths.  One
implication is that routing matrices have small {\em effective} rank
compared to their actual rank --- a small set of eigenvalues of $G^TG$
tend to be much larger than the rest.  Next, we propose a statistical
framework for approximating summary path metrics which can be used to
exploit the property of low effective rank of routing matrices.

The statistical framework we have developed draws on linear model
theory.  Linear model theory is concerned with the statistical inference of
quantities that are related through some underlying linear relationship.
In this case, we are concerned with the inference of path properties,
which are related through their linear dependence on link properties.
Linear model theory provides a well developed set of tools that allow one
in many cases to construct and analyze procedures for making
optimal linear inferences.

While our previous work has developed a framework for attacking the
problem, there are a number of hurdles, both theoretical and practical,
separating our simple model from application to realistic network
settings.  Our goal in this paper is to eliminate those hurdles.  We do
so by a combination of analysis, experimentation, and application to
real network data.  In the process, we demonstrate the power of the
resulting framework for solving problems of real interest
to network operators and users.

Our first contribution concerns the generality of our methods.  Our
initial, simple analysis relied on a model in which link metrics are
assumed to have equal variance.  This property does not hold in
practice; in reality, the variance of link delay (for example) can vary
considerably from link to link, depending on phenomena such as localized
congestion.  It is important to understand how the structure of our
proposed solution changes when link metrics have unequal variance, as
occurs in practice.  We first present the analysis of this more general
case and show how to incorporate this more realistic property into our
framework.

As previously mentioned, the opportunity for efficient network-wide
measurement derives from the small effective rank of routing matrices.
However the factors that determine this phenomenon are not fully
understood.  Thus it is important to ask whether the low effective rank
of routing matrices is a robust property.  This property can be affected
in at least two ways: first, if links fail, the degree of link sharing
among paths is affected; and second, if link metrics have unequal
variance, the benefits of link sharing may be reduced.  Our second
contribution is an evaluation of how effective rank changes when network
links fail, and when links have unequal variance.  We show that, even in
the presence of significant differences in variance across links,
unequal link variance does not generally diminish the utility of our
methods.

Since our goal is to move our methods into practical use, and to
demonstrate their utility, it is appropriate to apply them to data
obtained from a real network.   For that purpose we employ a large set
of measurements from the NLANR Active Measurement Project (AMP).  These
are comprised of per-hop delay measurements; we select the subset of
these measurements traversing the Abilene network.  Throughout the paper
we assess our analytic assumptions as well as the actual performance of
our methods using this data.

Our work is motivated by a number of current problems in network
measurement.  Many projects are currently using all-pairs path
measurement, including \cite{NLANR,RIPETT,ITR}.  To the extent that such
projects intend to either measure network-wide health or detect
anomalous network conditions, we hope that our methods can enable more
efficient measurement strategies.  With respect to measurements of
subnetworks, research such as \cite{akella04:multihome_performance} has
shown that significant performance benefits can accrue when a customer
chooses the best path from among a set of alternatives made available by
multihoming.

To demonstrate our applicability to these sorts of problems, we
evaluate the performance of our methods on Abilene data for a number
of specific cases.  The first problem is that of obtaining
network-wide averages of per-path delay, as would be needed for a
network ``health'' measure.  We show that a very accurate estimate of
average path delay (with relative error typically less than 1\%) is obtained
using as few as one-tenth of the paths needed for exact measurement.
In addition, we show that our extended
methods that incorporate unequal link variance are more accurate than
the simpler methods in our previous work.

Our second problem concerns the detection of anomalies within the
network.  We are interested in detecting when average network
performance exceeds a threshold, such as three standard deviations from
its mean.  We show that reasonable anomaly detection can be done 
with a
very small subset of networks paths --- as few as only 
one-third of the paths needed for exact measurement. 

Our final problem concerns the problem of selecting between two
network ingress points.  In this setting, a network customer or peer
has multiple connection points to the network of interest and needs to
choose between them for accessing a given set of destinations.  We
show that, once again, only a small set of paths need to be measured
in order to accurately choose between the alternatives.

The rest of the paper is organized as follows.  In Section~\ref{sec:bg}
we review previous work in this area and provide necessary background.
Then in Section~\ref{sec:methodology} we present our analytic results,
which extend our previous methods to the case of unequal link variance.
Next in Section~\ref{sec:data} we describe the data used in evaluating
our methods.  In Section~\ref{sec:redundancy} we evaluate the robustness
of low effective rank of routing matrices, and in
Section~\ref{sec:applications} we describe the results of applying our
methods to the three real-world problems. Finally in
\ref{sec:conclusion} we conclude.

\section{Background}
\label{sec:bg}
In this section we formally define the path estimation problem and discuss
previous work that has addressed it. 
Let $\G = (\V, \E)$ be a strongly connected directed graph, where the nodes
in $\mathcal{V}$ represent network devices and the edges in
$\mathcal{E}$ represent links between those devices. Additionally, let
$\mathcal{P}$ be the set of all paths in the network. Let $n_v =
\abs{\V}$, $n_e = \abs{\E}$, $n_p = \abs{\P}$ denote respectively the
number of devices, links and paths.

We will consider a metric measured on the paths $i\in\P$ whose value
$y\in\R^{n_p}$ is a linear function of the value $x\in\R^{n_e}$ of the
same metric on the edges $j\in\E$. In particular, we are interested in
the case where $n_p\gg n_e$ and the linear relation between $y$ and $x$
is given by $y=Gx$, where $G\in\{0,1\}^{n_p\times n_e}$ is a routing
matrix whose entries simply indicate the traversal of a given link by
a given path, so that
\begin{equation}
  \label{eq:1}
  G_{i,j} = 
  \begin{cases}
    1 & \text{if path $i$ traverses link $j$} \\
    0 & \text{otherwise.}
  \end{cases}
\end{equation}
For example, if we let $x$ denote the delay times for edges in the
network and let $y$ denote the delay times for paths in the network,
then $y=Gx$.  A similar relation holds, under suitable transformation,
between path-wise and link-wise loss rates.

As explained in Section~\ref{sec:introduction},
our interest in this paper focuses on the problem of monitoring global
network properties via measurements on some small subset of the paths.
Earlier work by Chen and colleagues \cite{chen03:overlay} shows that in fact
one need measure only a number of paths $k^*$ on the order of the number of
links in the network, to still be able to recover {\it exact} knowledge of all
network path behaviors.  Their argument is essentially linear algebraic in
nature, and is based upon the fact that a subset, say $\tilde G$, of only
$k^*=\rank(G)$ independent rows of $G$ are sufficient to span the image of $G$
\emph{i.e.}, $\{y\in\R^{n_p} : y=Gx, x\in\R^{n_e}\}$.  As a result, given the
measurements for paths corresponding to the rows in such a $\tilde G$,
measurements for all other paths may be obtained as a function thereof.
Similar work may be found in~\cite{nguyen04:link_failures}, in the context of
Boolean algebras, for the problem of detecting link failures.

Following on their work in~\cite{chen03:overlay}, Chen and colleagues
showed~\cite{chen04:algebraic} that the number $k^*$ of paths needed for their
method scales at worst like $O(n_v\log n_v)$ in a collection of real and
simulated networks.  Furthermore, they argue that this behavior is
to be expected in internet networks, due to the high degree of sharing
of paths as they traverse common routes in the dense core.  

In our previous work~\cite{blind}, such sharing was noted as well.  However,
there it was shown that the {\it effective} rank of $G$ can in fact be
arguably much less than the actual rank.  As a result, we proposed
that substantially greater savings can be achieved by allowing for the {\it
  approximate} monitoring of path metrics and adopting an approach based on
predictive linear statistical methods.

\begin{figure}[tbp]
  \centering
  \subfigure[Map of Abilene]{%
    \label{fig:abilene-map}
    \includegraphics[width=2.5in]{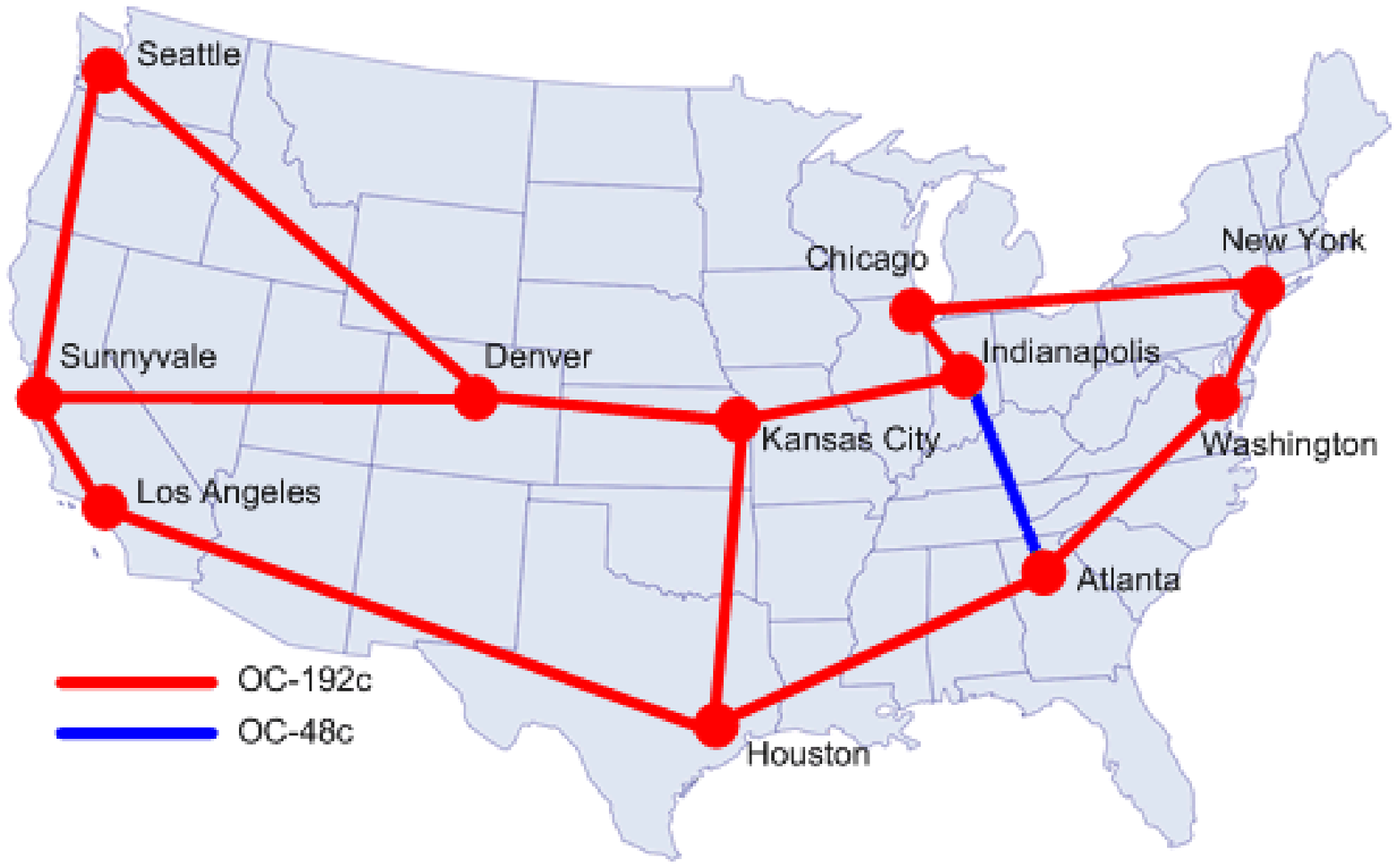}}
  \quad
  \subfigure[Eigenspectrum of $G$]{%
    \label{fig:abilene-spec}
    \includegraphics{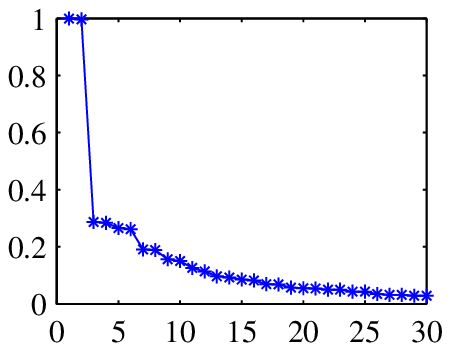}}
  \caption{Map of the Abilene network and the eigenspectrum of one of its
           routing matrices.}
  \label{fig:map-and-spec}
\end{figure}

As an illustration, consider the Abilene network,
a high-performance network which
serves Internet2 (the U.S.\ national research and
education backbone).  A map of this network is shown in 
Figure~\ref{fig:abilene-map}.  The network can be seen to consist 
of $11$ nodes, at locations across the continental United States, but only 
$2\times 15=30$ directed links.  Accordingly, a large amount of sharing
of these links can be expected of the $11\times 10=110$
paths on the network.

Our work goes beyond these observations to point out that in fact, some links
are shared much more highly than others. The effect can be evaluated via the
eigenspectrum of the routing matrix. 
In Figure~\ref{fig:abilene-spec} is shown the eigenspectrum 
of a routing matrix $G$ for this network.  More specifically, we
have plotted the (ordered) eigenvalues of the symmetric, 
non-negative definite matrix $G^TG$.  Since all $30$ of these
eigenvalues are positive, we have that $\rank(G)=30$, and therefore,
recalling the results in~\cite{chen03:overlay,chen04:algebraic},
no more than $30$ paths need to be measured in order to recover
exact knowledge of an additive metric $y$.  However, the decay
in the spectrum in Figure~\ref{fig:abilene-spec} --- particularly
the sharp decay in the early portion --- suggests that the effective
rank of $G$ may be a good deal less than 30, perhaps on the order of
10 or so.  Algebraically, this means that the span of the rows of $G$
exists primarily (but not entirely)
in a subspace of $\R^{n_p}$ of dimension potentially much lower
than 30.  From a practical perspective, this means that accurate
(but not exact) recovery of $y$ may be possible based on measurements
from much fewer than 30 paths.

We have confirmed this behavior of the routing matrix $G$ 
in additional real networks of varying and substantial sizes;
and we have exploited the decay in these matrices in developing a 
framework for efficient monitoring of paths
based on statistical prediction, in which measurements from measured
paths are used to predict those on unmeasured paths~\cite{blind}.
However, this earlier work was developed in the context of a relatively
simple model, in which link measurements were assumed to be 
uncorrelated and to share a single, common variance.  Additionally,
the resulting methodology was only validated through 
analytics and numerical simulations.  Nevertheless, these validations
suggested a strong potential for the underlying concept; here we show
that that potential is realizable.

\section{Efficient Monitoring}
\label{sec:methodology}

In this section we describe a linear statistical framework for prediction
of end-to-end properties. This is a general framework, which includes
our previous work as a special case and is more
immediately applicable to real-life data.  Additionally, it uses a 
correspondingly different criterion for selecting 
which paths to measure, one that incorporates
information on both path sharing and relative variability of
the metric on links. The impact of these
changes is not only analytic or abstract; there can be a substantial
effect on the selection of paths, as we illustrate in
Section~\ref{sec:effect-of-link-cov}. 

\subsection{Statistical Prediction}
\label{sec:stat-pred}

We begin by building a model for the end-to-end properties in $y$.
In the work that follows, it is necessary only that the 
first two moments of $x$ and $y$ be specified, as opposed
to a full distributional specification.
Let $\mu$ be the
mean of $x$ and let $\Sigma$ to be the covariance of $x$.  Then the
corresponding statistics for $y$ are simply $\nu=G \mu$ and $V = G
\Sigma G^T$, respectively.

Now fix $k \leq \rank(G)$.  Let $y_s\in\R^k$ denote the values
$y_{i_1},\dots, y_{i_k}$ of the metric of interest for $k$ paths
$i_1,\dots,i_k\in\P$ that are to be sampled (\emph{i.e.}, measured),
and let $y_r\in\R^{n_p-k}$ denote the values for those $n_p-k$ paths that
remain.  Similarly, let $G_s$ be those rows of $G$ corresponding to the $k$
paths, $i_1,\dots,i_k$ and let $G_r$ be the remaining rows.  Then 
we may partition $y$ and $G$ into
\begin{equation}
  \label{eq:2}
   y =
  \begin{bmatrix}
    y_s \\ y_r
  \end{bmatrix}
  \text{\quad{}and\quad}
  G =
  \begin{bmatrix}
    \Gs \\ \Gr
  \end{bmatrix}
  \text{,}
\end{equation}
and we may similarly re-express the mean and covariance of $y$ as
\begin{equation}
  \label{eq:3}
  \nu = 
  \begin{bmatrix}
    \nu_s \\ \nu_r
  \end{bmatrix}
  = 
  \begin{bmatrix}
    G_s \mu \\ G_r \mu
  \end{bmatrix}
\end{equation}
and
\begin{equation}
  \label{eq:4}
  V = 
  \begin{bmatrix}
    \Vss & \Vsr \\
    \Vrs & \Vrr
  \end{bmatrix}
  = 
  \begin{bmatrix}
    \Gs \Sigma \Gs^T & \Gs \Sigma \Gr^T \\
    \Gr \Sigma \Gs^T & \Gr \Sigma \Gr^T
  \end{bmatrix}
  \text{.}
\end{equation}

In this paper, we take as our monitoring goal the task of obtaining
accurate (approximate) knowledge of a linear summary
of network path conditions. That is, we seek to accurately predict
a linear function of the path conditions $y$, of the form $l^Ty$
where $l\in \R^{n_p}$, based on the measurements in $y_s$. 
Two such linear summaries are the network-wide average, given 
by $l^T y$ for $l_i \equiv 1/n_p$,
and the difference between two groups
of paths $\P_1$ and $\P_2$, given by 
\begin{equation}
  \label{eq:5}
  l_i = 
  \begin{cases}
    1/\abs{\P_1} & \text{if $i \in \P_1$} \\
    -1/\abs{\P_2} & \text{if $i \in \P_2$} \\
    0 & \text{otherwise.}
  \end{cases}
\end{equation}
The prediction of $l^Ty$, from the $k$ sampled path
values in $y_s$, can be viewed as a
particular instance of the classical problem of prediction in the
statistical literature on sampling
\cite{valian00:finite_pop_sampling}.

If the mean-squared prediction error (MSPE) is used to judge the 
quality of a predictor \emph{i.e.}, if the quality of a predictor $p(y_s)$ 
is measured by $E[(l^Ty - p(y_s))^2]$,
then the best predictor is known to be given by the conditional
expectation
\begin{equation}
  \label{eq:6}
  E[l^Ty|y_s] = l_s^T y_s + E[l_r^Ty_r|y_s],
\end{equation}
where $l_s$ and $l_r$ are defined similarly 
to $y_s$ and $y_r$, respectively. 
But this requires knowledge of the joint distributional structure. It
is therefore common practice to restrict attention to a smaller and
simpler subclass of predictors, with the class of linear predictors being a
natural choice.  In this case, the best linear predictor (BLP) is given by
the expression
\begin{equation}
  \label{eq:7}
  a^T y_s = l_s^T y_s + l_r^T \Gr \mu + l_r^T c_* (y_s - \Gs \mu) \text{,}
\end{equation}
where $c_*$ is any solution to $c_* \Vss = \Vrs$. (The derivation of
this result follows similarly to the analogous result for simple
linear statistical models given in \cite[VI.3]{christensen87:plane}.)

However, without knowledge of $\mu$, the BLP in (\ref{eq:7})
is an ideal that cannot be
computed.  One natural solution is to estimate $\mu$ from the data.
This is a version of the network tomography problem, and in some
sense the reverse of the well-known traffic-matrix estimation
problem. There are many versions of this problem, and many methods for
solving them.  See~\cite{CoatesEtAl02}, for example.  We will use a
simple method based on generalized 
least squares to estimate the mean $\mu$ as
\begin{equation}
  \label{eq:8}
  \muhat = [\Gs^T \Vss^{-1} \Gs]^{-} \Gs^T \Vss^{-1} y_s\text{.}
\end{equation}
Here $M^{-}$ denotes a generalized inverse of the matrix $M$. 

Substituting $\muhat$ for $\mu$ in (\ref{eq:7}) produces an estimate
of the BLP (an E-BLP) that is a function of only the
measurements $y_s$, the routing matrix $G$ and the link covariance
matrix $\Sigma$.  Specifically, we obtain
\begin{equation}
  \label{eq:9}
  \begin{split}
    \ahat^T y_s &= l_s^T y_s + l_r^T \Gr [\Gs^T \Vss^{-1} \Gs]^{-} \Gs^T
    \Vss^{-1} y_s \\
    &= l_s^T y_s  + l_r^T \Vrs \Vss^{-1} y_s \text{.}
  \end{split}
\end{equation}
The first expression in (\ref{eq:9}) can be reduced to the second  
using properties of generalized inverses and projection matrices.
The derivation of these and the other expressions above requires only that
$\Vss$ be invertible and that $\Sigma$ be positive definite.

As a side point, it is useful to note that
the form of our E-BLP also follows from application of the 
arguments in~\cite{blind} to the transformed problem $y = (GC) (C^{-1}
x)\equiv\tilde{G} \tilde{x}$, where $C$ is a non-singular matrix
deriving from the factorization $\Sigma=CC^T$.  The effect of
this transformation is to introduce the variable
$\tilde x$, with mean $C^{-1}\mu$ and covariance $I$, whose simplified
covariance structure then follows that assumed for the calculations
in~\cite{blind}.  

\subsection{Path Selection}
\label{sec:subset-selection}

The material in Section~\ref{sec:stat-pred} assumes a set of measurements
from $k$ paths $i_1,\ldots,i_k\in\mathcal{P}$.  However, 
given the resources to measure any $k$ paths in a network, we are still
faced with the question of which $k$ paths to measure.  
A natural response would be to choose $k$
paths that minimize $\mspe(\ahat^T y_s)$,
over all subsets of $k$ paths.  
This quantity can be shown to have the form
\begin{equation}
  \label{eq:10}
  \begin{split}
    \mspe(\ahat^T y_s) &= \mspe(a^T y_s) + (\bias \ahat^T y_s)^2
  \end{split}
\end{equation}
where
\begin{equation}
  \label{eq:11}
    \mspe(a^T y_s)
    = l_r^T \left( \Vrr -  \Vrs \Vss^{-1} \Vsr\right) l_r 
\end{equation}
and
\begin{equation}
  \label{eq:12}
    \bias(\ahat^T y_s) 
    = l_r^T \left(\Vrs \Vss^{-1} G_s - \Gr\right) \mu
  \text{.}
\end{equation}
We see that the error inherent in our E-BLP, $\hat a^Ty_s$,
is equal to that of the BLP, $a^T y_s$, which is an unbiased
predictor, plus the bias that accompanies the need to estimate
$\mu$.  Accordingly, the expression for $\bias(\ahat^T y_s)$
involves both the mean $\mu$ and the covariance $\Sigma$ for the
unseen link measurements $x$, whereas that for $\mspe(a^T y_s)$
involves only the covariance.  

Of course, since we typically do not have knowledge of $\mu$,
minimization of the full expression for $\mspe(\ahat^T y_s)$
in (\ref{eq:10}) is an unrealistic goal in practice.  Instead,
if adequate information on the covariance matrix $\Sigma$ is available,
one might consider trying to minimize the expression
for $\mspe(a^T y_s)$ in (\ref{eq:11}).  
A useful equivalent expression for this MSPE is
\begin{equation}
  \label{eq:13}
    \mspe(a^T y_s) 
    = l_r^T (\Gr C)
    (I - B_s)
    (\Gr C)^T l_r
        \text{,}
\end{equation}
where
\begin{equation}
  \label{eq:14}
  \begin{split}
    B_s &= (\Gs C)^T [(\Gs C)(\Gs C)^T]^{-1} (\Gs C) \\
    &= C^T \Gs^T\Vss^{-1} \Gs C
  \end{split}
\end{equation}
is the orthogonal projection onto $\row(\Gs C)$.  Since orthogonal
projection matrices are idempotent and symmetric, the MSPE in
(\ref{eq:13}) can be seen to be the 
square of the Euclidean norm of the projection
of $(\Gr C)^T l_r$ onto $\row(\Gs C)^\perp = \nul(\Gs C)$. 

The nature of the space $\row(\Gs C)$ obviously will depend on 
$G$, $C$, and the manner in which they interact through multiplication
of the latter by the former.  Here, motivated by
our examination of the covariance structure of the edge delay 
data in Section~\ref{sec:data}, we consider in detail the case
where $\Sigma$ is a diagonal matrix.  We leave the problem of
path selection for non-diagonal covariance as an interesting open 
problem.

To begin, suppose that $\Sigma = I$ and
consider the case of predicting the metric on a single path that lies
outside of the sample ($l_s$ all zeros and $l_r$ containing a single
one). In this case we have $C=I$, so the MSPE simply measures the
degree to which the row of $G$ corresponding to the single path lies
outside of $\row(G_s)$.  Of course, if our goal is only to monitor
the value $y_i$ for a single path $i\in\mathcal{P}$, then
the most efficient strategy would be simply to measure $y_i$.
However, we are interested here in strategies that allow for
the capacity to do whole-network monitoring, in which case the 
minimization of (\ref{eq:13}), when $C=I$ and $l$ is nontrivial, 
can be interpreted roughly as
seeking a subset of $k$ paths for whom the rows in $G_s$ capture as 
many of the rows of $G$ as possible to the largest extent possible.
This situation corresponds to that studied
in~\cite{blind}.

Now take the case where $\Sigma$ is an arbitrary diagonal matrix.  
The matrix $C$ is
then also diagonal and the columns of $GC$, which correspond to links in
the network, are the columns of $G$ rescaled by the variability of the
corresponding link.  That is, each non-zero entry in a given row
of $G$, indicating that the corresponding path passes over a given link,
is replaced by the standard deviation of the metric on that link.
Therefore, we see that
minimization of the MSPE in the case of diagonal $\Sigma$
is analogous to the case of $\Sigma=I$, but with the important 
difference that the routing information for each path is augmented
with information on the variability among the links over 
which it travels, thus allowing for the relative weight among these
links to differ.  We will see in Section~\ref{sec:effect-of-link-cov}
that the incorporation of such variance information can have a 
nontrivial effect on which paths are selected.

From the standpoint of optimization theory, our path-selection problem
may be viewed as an example of the so-called `subset selection'
problem in computational linear algebra. 
In the two cases just described, the selection of an appropriate subset
of rows of $GC$ has a meaningful physical interpretation, in terms of
the selection of paths, and vice versa.
Exact solutions to this problem are computationally infeasible 
(it is known to be NP-complete), 
but the problem is well-studied and an assortment of
methods for calculating approximate solutions have been offered.  

The method we have used for the empirical work in this paper was
adapted from the subset selection method described in Algorithm
12.1.1 of \cite{golub83:matrix_comp}\footnote{The reader is referred
  the standard reference \cite[Chap.\ 12]{golub83:matrix_comp} for
  further details.}, and is a modified version of that described
in our previous work~\cite{blind}. 
Specifically, we first compute a singular value decomposition 
$GC=U\Delta V^T$ for $GC$.  The columns of the matrix $U$ form 
an orthogonal basis for the span of $GC$, with the relative importance
of each column indicated by the magnitude of the corresponding singular
value in the diagonal matrix $\Delta$.  These values in turn are
simply the square-root of the eigenvalues of $(GC)^T(GC)$, through
which the connection with the discussion on reduced-rank 
in Section~\ref{sec:bg} is evident.  Ideally,
we seek a subset of $k$ paths for which the corresponding rows of $GC$
span the first $k$ singular dimensions.  We make heuristic use of a
QR-factorization with column pivoting to pursue this task,
writing $U_{(k)}^T P_k = QR$, where $U_{(k)}$ is an $n_p\times k$ matrix
formed from the first $k$ columns of $U$ and $P_k$ is an $n_p
\times n_p$ permutation matrix defined by the column pivoting. We then
take $\Gs$ to be the submatrix of $G$ formed by the first $k$ rows of
$P_k^T G$.

It should be noted that $P_k$ has to be recomputed for different
values of $k$.  However, the overall complexity for the computation of
E-BLP in (\ref{eq:9})
is dominated by the computation of the SVD of $GC$ which is $O(n_p^2
n_e)$.  This can likely be improved through the use
of methods for sparse matrices, since the entries of $GC$ tend to
include a large fraction of zeros.
The other components are the QR-factorization with column
pivoting which is $O(k^2 n_p)$ and the computation of $\Vss^{-1}$
which is only $O(k^3)$.

\section{Data: Abilene Path Delays}
\label{sec:data}

Our methods are applicable to any per-link metric that adds to
form per-path metrics.  To demonstrate a concrete example we take
per-link delays for Abilene.

\subsection{Collection and Processing}
\label{sec:data-processing}

Our data comes from the NLANR Active Measurement Project.  This project
continually performs traceroutes between all pairs of AMP monitors on
ten minute intervals.  Because most AMP monitors are on networks with
Abilene connections, most traceroutes pass over Abilene.  This provides
a highly detailed view of the state of the Abilene network.  Our data
consists of the set of all measurements taken over 3 days in 2003,
yielding 432 time points.

For each time period, we start with traceroutes between the 14917 pairs
of monitors for which complete data was available.  From this set we
construct estimates of a consistent set of delays for each link within
the Abilene network.  Links comprising the Abilene network were
identified by their known interface addresses.  Since different
traceroutes traverse each link at slightly different times, and since
each traceroute takes up to three measurements per hop, we form a single
estimate of each link's
delay by averaging across all the traceroutes that measured that
link in the current time point.   This yields a single measure of delay
for each link and timepoint;  while this measure does not capture the
variations in delay that occur within a ten minute interval, it provides
a realistic and representative value for delay that can be used to
construct path-wise metrics.

Note that in practice, our methods do not make use of per-link
measurements; we only need to form them for
validation purposes.   Thus, issues with accurate measurement of
per-link delays do not represent a limitation of our methods.

\subsection{Data Characteristics}
\label{sec:data-characteristics}

In our data, the mean delays for Abilene's $30$ directed edges are evenly
spread out from 2 to 36 milliseconds, with standard deviations that run from
0.16 to 0.94 for the full three-day period. 
Other than a diurnal cycle in some of the edges, we found nothing
remarkable in the structure of the edge delays.  There is no apparent
relationship between the edge delays and the standard deviations;
edges with similar means exhibit both large and small standard
deviations. 

In order to learn more about the 
relationships between the edge delays
themselves, we looked at the covariance matrix for one day's
worth of 
data. 
The entries in this
matrix are primarily dominated by the diagonal elements, with a small
number of off-diagonal entries of similar magnitude.
However, inspection of the actual delay data on the links involved
in the large off-diagonals leads us to believe that these cases are actually
artifacts of the measurement procedure. 

Note that our framework in Section~\ref{sec:methodology} requires
knowledge of the covariance matrix, which in practice could be
obtained from either historical data or possibly periodic, infrequent
measurement of the edges.  For the purposes of this paper, and
motivated by the findings mentioned just above, we will take 
$\Sigma$ to be a diagonal covariance matrix with non-zero entries
given by the variances of the edge delay data over one day.

\section{Robustness of Path Redundancy}
\label{sec:redundancy}

Critical to the success of our proposed methodology is the degree to
which some subset of independent rows in $GC$, denoted $\Gs C$ above,
can effectively approximate the span of the full set of rows in $GC$.
Since multiplication by a diagonal $C$ does not change the linear
relations between the rows of $G$, this boils down to an issue of path
redundancy.  
In order for path redundancy to translate into reduced measurement frameworks
with dependable implementation and performance characteristics,
it needs to be a robust property.  We
examine that robustness here by looking at the effect of two factors:
(1) inclusion of information on link variances, via the matrix $C$, 
and (2) network integrity.
We find the redundancy in the cases we study to be quite robust.

\subsection{Effect of Link Covariance}
\label{sec:effect-of-link-cov}
The path selection scheme proposed in this paper takes into account not 
only the sharing among paths, but also the relative variances of the link
measurements.  Therefore, it is natural to ask to what 
degree inclusion of the information on the link measurement variances
affects the relative rank involved.
Analysis of the Abilene data provides evidence to suggest that the
phenomenon of reduced rank should remain robust
to the incorporation of link variance information
that varies across a moderate range, while that information
in turn can lead to meaningful adjustments in the finer details of the path
selection process.

\begin{figure}[tbp]
  \centering
  \includegraphics{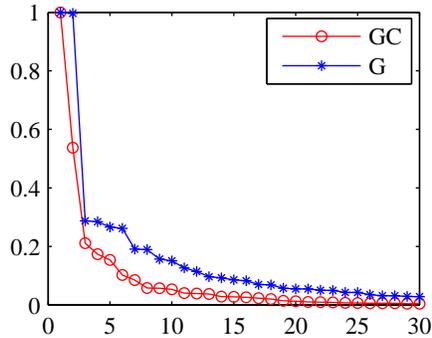}
  \caption{Comparison of eigenspectra for $G$ and $GC$ on Abilene.}
  \label{fig:GvGCspecs}
\end{figure}

Our analysis is based primarily on a comparison of the behavior
of the eigenvalues and eigenvectors of the matrix $(GC)^T(GC)$,
in the two cases of $\Sigma=I$ and $\Sigma$ the diagonal matrix
described in Section~\ref{sec:data-characteristics}.  
The eigenspectra of $GC$ for these two cases, shown in
Figure~\ref{fig:GvGCspecs}, are quite similar. 
In particular, they exhibit similar
decay, which suggests a corresponding similarity in the relative ranks
of $G$ and $GC$.  In turn, we may therefore expect similar relative performance
for a given number of measured paths.  This expectation is found to be
fulfilled in the application presented in 
Section~\ref{sec:network-wide-average}.

\begin{figure}[tbp]
  \centering
    \includegraphics{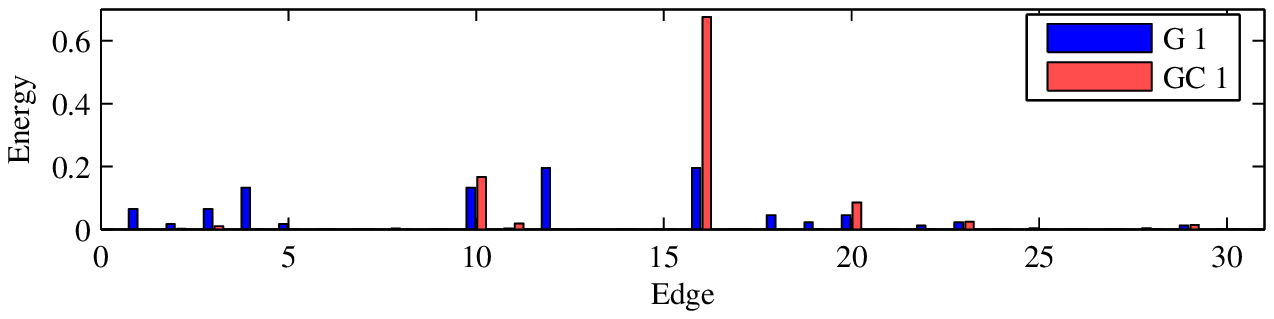}

    \includegraphics{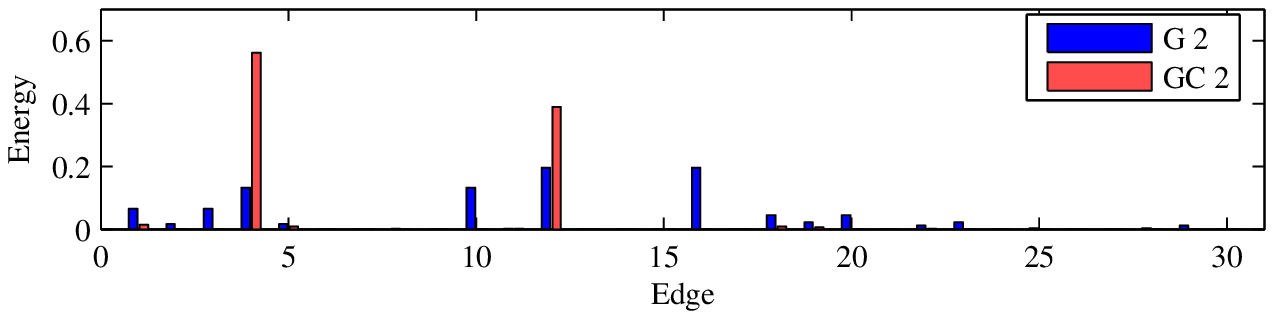}
  \caption{Comparison of the energy distribution of first two 
    eigenvectors of $G^TG$ and $(GC)^T(GC)$. Top Figure: 
    First eigenvector
    of $G^TG$ versus the first eigenvector vector of $(GC)^T(GC)$.  
    Bottom Figure: Second eigenvector
    of $G^TG$ versus the second eigenvector of $(GC)^T(GC)$. 
    (See Appendix~\ref{sec:abilene-edges} for a list of edges.)} 
  \label{fig:first-two-sing-vec-G-vs-GC}
\end{figure}

However, despite their similarity, the two spectra are not identical,
and it is interesting to explore further the reasons for and the
implications of their differences.  For example, note that in the
spectra for $G^TG$ there appears to be a pattern of pairing among the
larger eigenvalues, whereas this pattern is absent from the spectra of
$(GC)^T(GC)$ in the case of $\Sigma$ diagonal.  The pairing in the
case of the former derives from the fact that the routing is symmetric
under the particular routing matrix $G$ we obtained.  The lack of
pairing in the case of the latter is a reflection of the unequal link
variances.  This observation is supported by an examination of the
first two eigenvectors
in our two models, shown in
Figure~\ref{fig:first-two-sing-vec-G-vs-GC}, in conjunction with the
Abilene map, which reveals that in both cases their energy is
concentrated on links making up both directions of the northern
transcontinental route---particularly in the region of Indianapolis.  But
an examination of the variances in our diagonal model shows that the
edges along the eastbound route, in particular 
Kansas
City--Indianapolis (edge 16), generally have larger variances than
their westbound counterparts.  Hence the gap between the first and
second eigenvalues in this model, and the concentration of the energy of
the first 
eigenvector of $(GC)^T (GC)$ on the edges of the eastbound route. 

\begin{figure*}[tbp]
  \centering
  \includegraphics{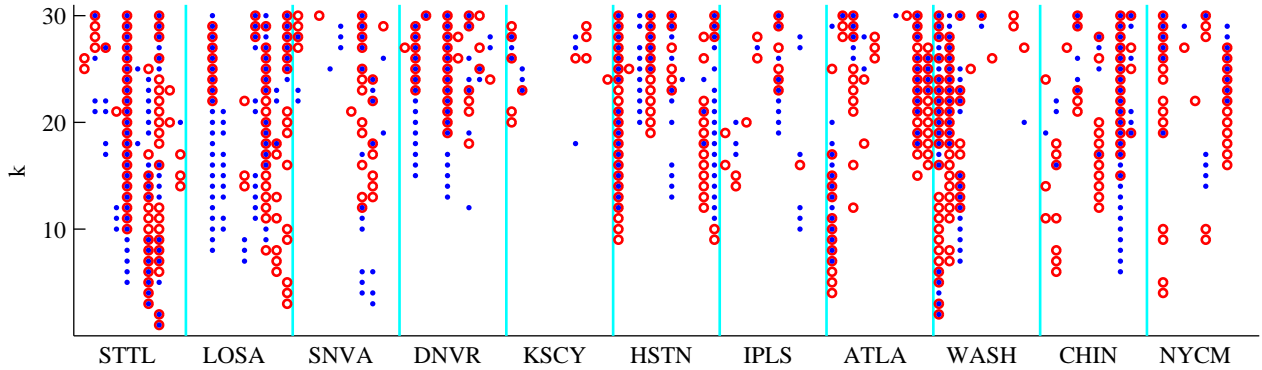}
  \caption{Subset selection for $\Sigma = I$ compared to selection for
    diagonal $\Sigma$. The markers in each row represent the paths
    chosen for the size $k$ sample when $\Sigma = I$ (red circles) and
    when $\Sigma$ is diagonal (blue dots).}
  \label{fig:GC-subset-sel}
\end{figure*}

Further insight into the effect of link variance information can be
obtained by looking at the actual paths selected by our algorithm under
each choice of $\Sigma$.  These are shown in
Figure~\ref{fig:GC-subset-sel}.   Each column corresponds to a 
path, and the markers in a given row $k$ indicate which $k$ paths are selected.
The paths have been grouped according to their starting node;
the ten paths within
each group are sorted by destination using the same ordering.
So for example, the first column in the Chicago group
corresponds to the Chicago--Seattle path.

A number of interesting characteristics may be observed.  
A large fraction of the chosen paths are found to match,
particularly for larger $k$.  However, under the diagonal $\Sigma$
there does seem to be a slightly increased emphasis
on eastward paths, as might be expected from our earlier discussion.
When the paths do differ, often the change in paths will be minor,
in the sense that the ingress point remains the same and the egress
point changes to a neighboring node (red circles next to blue dots).
Similarly, other times the ingress
point will move to a geographical ``neighbor''. 
This behavior is especially apparent in the case of smaller values
of $k$; for example in $k=3$ to $6$ the eastward
paths from  Los Angeles are traded for eastward paths starting
in Sunnyvale. 

\subsection{Effect of Link Failures}
\label{sec:link-failures}

The analysis just described is for the Abilene routing matrix
when the network is fully intact.  Failures of links within a 
network are to be expected, however, and our method should be 
robust to such failures.  In particular, it is to be hoped that
the reduced rank of $GC$ persists sufficiently under
the failure of a link(s) to allow for the continued use of a (possibly 
altered) reduced subset of paths for monitoring.  An analysis of
our Abilene data suggests that a substantial degree of
robustness can be expected for path redundancy
in the face of normal link failures.  That is, both the number of
paths needed to capture a given portion of the span of $GC$ and
the actual choice of paths appeared quite stable.

\begin{figure}[tbp]
  \centering
  \subfigure[]{%
    \includegraphics[width=\twofig]{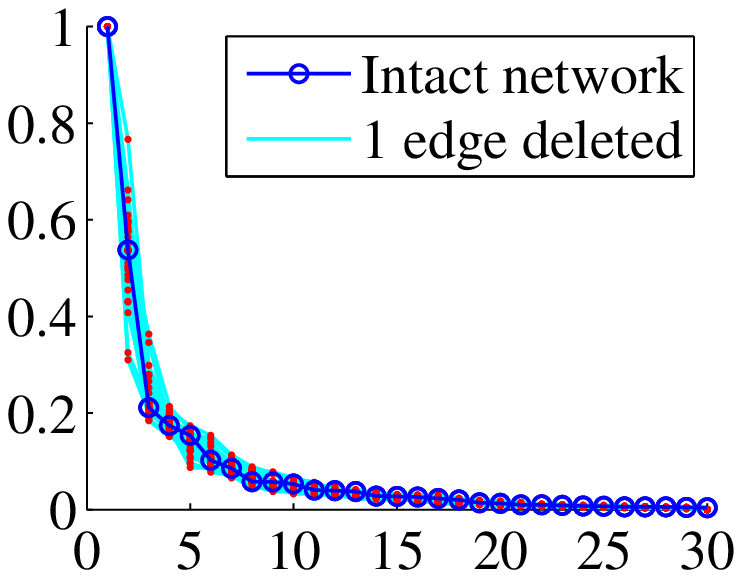}
    \label{fig:delSpec1-scaled}}
  \qquad
  \subfigure[]{%
    \label{fig:delSpec2-scaled}
    \includegraphics[width=\twofig]{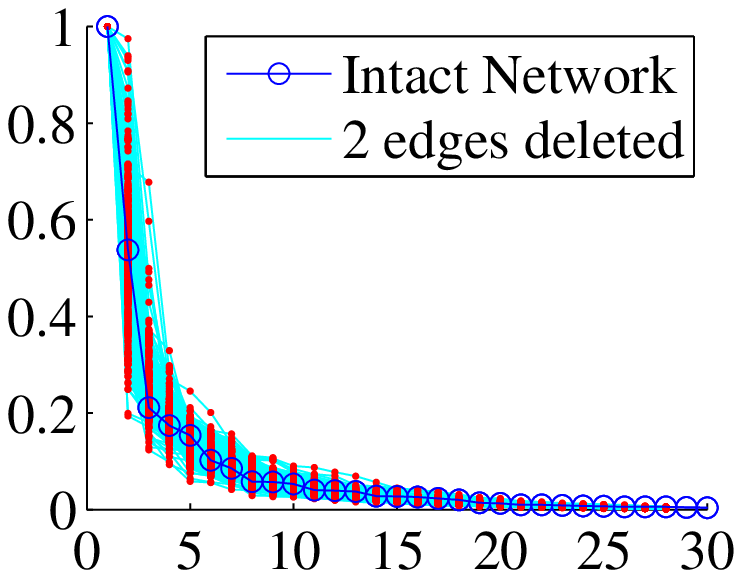}}
  \caption{Spectra for $GC$ of Abilene delay data,
        after the deletion of (a) one or (b) two links. 
    In Figure (b) the spectra for the 18
    graphs that are not strongly connected are not displayed. Each
    spectrum has been rescaled so that the first eigenvalue is one.}
  \label{fig:delSpec-1-and-2-scaled}
\end{figure}

Proceeding
as in Section~\ref{sec:effect-of-link-cov}, we pursued these issues
by examining the behavior of the eigenvalues and eigenvectors of $GC$.
First consider the case of a single link failure.  Starting with 
a set of IGP weights for Abilene, we deleted one link and used
Dijkstra's algorithm to compute the shortest paths between cities.
These paths were then used to construct a routing matrix $G$ for the
modified network.  The spectrum of $GC$ for each of the 30 resulting
routing matrices, scaled to all have maximum eigenvalue $1$, are plotted in
Figure~\ref{fig:delSpec1-scaled}. Impressively, the spectra
appear to be quite stable.  A similar plot is shown in
Figure~\ref{fig:delSpec2-scaled} for the case of two link deletions.
While the stability of the spectra is less than in the previous
case, it is still substantial when one
considers that Abilene is a small network with only $30$
links.
Note that, in the absence of real data for each of the
modified networks, we have used a diagonal $\Sigma$ for the intact Abilene
network throughout this section.

\begin{figure}[tbp]
  \centering
  \includegraphics{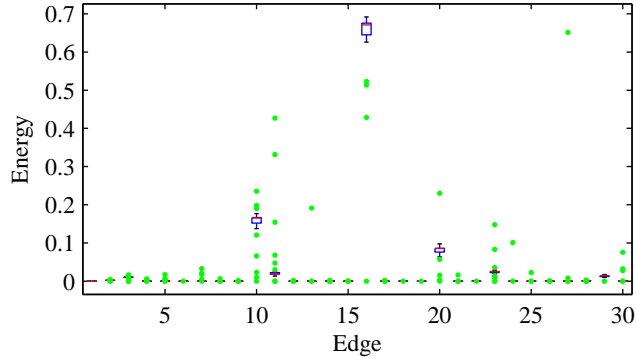}
  \caption{Boxplots of the energy distribution in the first
  eigenvector of $(GC)^T(GC)$ after one link deletion. The red bars
  represent the median and the green dots
  represent outliers. Note that many boxes have no height and are hidden
  behind the outliers.}
  \label{fig:delSpecEnergyDistrib}
\end{figure}

The stability we have observed in the spectra of
Figure~\ref{fig:delSpec-1-and-2-scaled} indicates that
even if a link or two goes down in the Abilene network, the number of paths 
one needs to measure for a given level of accuracy remains 
essentially unchanged.   But this conclusion still leaves unanswered
the question of \emph{which} paths should be used. To address this
question, we examine the stability of the energy distribution of the first 
eigenvector of $(GC)^T(GC)$.
Specifically, in Figure~\ref{fig:delSpecEnergyDistrib} we present the 
corresponding energy distributions following the
deletion of one link, for each of the 30 links, as described above.
The boxplots indicate that, generally, 
the energy distribution is quite stable under
link deletions; each box is centered tightly along
the energy distribution of the first eigenvector for the intact
network.

There are, however, a certain number of outliers, indicating that
certain particular edge deletions can have a notable effect.  For
example, one such outlier occurs when we delete edge 16 (Kansas
City--Indianapolis), which is the highest-energy  link for the intact
network.  
In this instance, the bulk of the energy is moved to the link
between Los Angeles and Houston, followed by a smaller allotment to
that between Houston and Atlanta.  This change would seem to indicate
a shift from the northern transcontinental route to the southern one.
Conversely, deletion of the link between Indianapolis and Chicago
(edge 10) causes a good portion of the energy to shift to the link
between Indianapolis and Atlanta.  These observations suggest that
a shift to the southern route occurs only when absolutely necessary.

\section{Applications}
\label{sec:applications}

In this section, we show how our framework may
be applied to address three practical problems of interest to
network providers and customers. 
In particular, we show how the
appropriate selection of small 
sets of path measurements can be used to accurately estimate
network-wide averages of path delays, to reliably detect network
anomalies, and to effectively make a choice between alternative
sub-networks, as a customer choosing between two providers or two
ingress points into a provider network.

\subsection{Monitoring a Network-wide Average}
\label{sec:network-wide-average}

An average is perhaps the most basic network-wide quantity that one
might be interested in monitoring. So as our first application, we
consider the prediction of the average delay over all paths. 

\begin{figure}[tbp]
  \centering
  \includegraphics{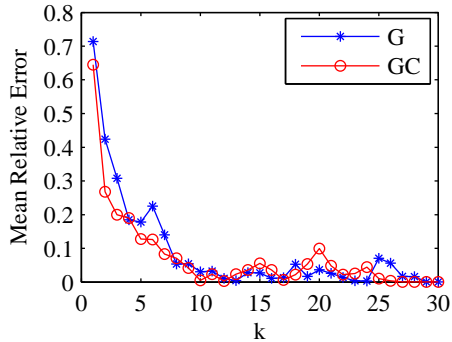}
  \caption{Mean relative prediction error as a function of $k$.} 
  \label{fig:rel-error}
\end{figure}

Recall from Section~\ref{sec:data} that our data consist of 
delays on all $n_p=110$ paths of the Abilene network, for each of 432
ten-minute epochs over three consecutive days.  Using (\ref{eq:9}),
with $l_i \equiv 1/n_p$, we computed predictions of the network-wide
average path delay during each epoch, for a choice of $k=1,\ldots,30$
measured paths, where the paths were chosen using the algorithm
described in Section~\ref{sec:subset-selection}.  Both the case of
$\Sigma=I$ and the case of a general diagonal $\Sigma$ were examined,
where the values for the latter derive from the analysis
described in Section~\ref{sec:data-characteristics}.
To summarize the accuracy of our predictions, we calculated the
average relative error for each $k$, where the average is taken 
over epochs.  The results are shown in Figure~\ref{fig:rel-error}.

Two conclusions are particularly apparent.  First, we see that
for both variance models increasing the number $k$ of paths measured
improves the accuracy of the prediction up
until around $k=9$ or $10$, after which it basically levels out.  
Since it is roughly at this point that the spectra of the (weighted)
routing matrices level out as well, this suggests that our subset
selection algorithm is indeed doing what we are asking of it, in that
it is tracking the effective rank quite closely.  Second, we note
that the method based on general diagonal $\Sigma$ typically outperforms 
that based on the more restricted assumption of $\Sigma=I$, again up until 
around $k=9$ or $10$, after which the relative performances oscillate in a
fairly random manner.  This indicates explicitly the gain that
can be gotten by incorporating more accurate variance information 
into the model where, for example, the relative error in the
more general model is on the order of just $5\%$ by
$k=9$. 

\begin{figure}[tbp]
  \centering
  \includegraphics{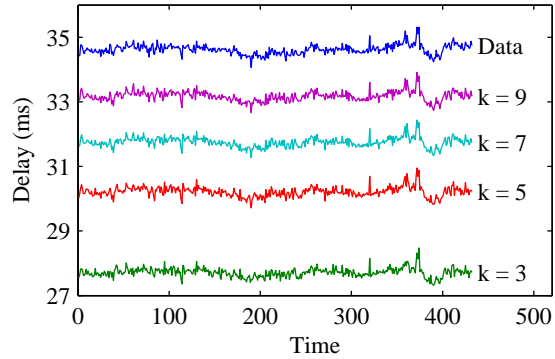}
  \caption{Predictions of network-wide average path delays, for
           various choices of $k$.}
  \label{fig:GC-preds}
\end{figure}

To get a better idea of how well the various predictors are doing, we
can compare plots of the predictions against a plot of the actual mean
delays, as shown in Figure~\ref{fig:GC-preds} for $k=3,5,7,$ and $9$.
Note that all of the predictions mirror the rise and fall of the actual
network-wide delay quite closely --- even for as few as $k=3$ measured
paths, for which the correlation between the two time-series is 
$\rho=0.814$.  However, it also is clear that there is a downward bias in
these predictions, and that this bias is increasingly prominent 
as $k$ decreases.  Therefore, some manner of bias correction is
needed.

The source of this bias can be traced to a lack of information 
on links in the network that end up being traversed by none of the
$k$ measured paths.  In fact, the generalized
inverse used in our E-BLP in (\ref{eq:9}) simply estimates the
corresponding values $x_j$ on these links to be zero.
Hence, as the measured paths reach a point where
every link contributes to at least one path, as it does by roughly
$k=10$, the bias diminishes accordingly.  Note, however, that the
bias for each $k$ in Figure~\ref{fig:GC-preds} is fairly constant.
This suggests that a small amount of additional measurement 
information could go a long way.

\begin{figure}[tbp]
  \centering
  \includegraphics{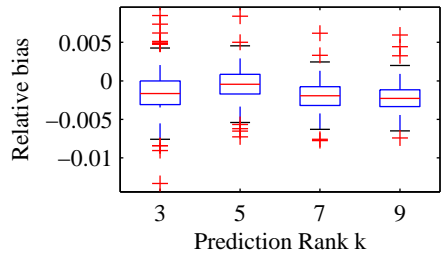}
  \caption{Relative bias after bias correction.}
  \label{fig:bias-corrected-ts}
\end{figure}

We implemented a simple method of bias correction, in which we 
assume access to full link measurements $x$ for the first of 
our 432 epochs.  This corresponds to a one-time use of path measurements among
a sufficient set of paths for complete reconstruction of link delays (in this
case, 30 paths). Since it is a one-time-only measurement, it represents a
minimal addition to the network measurement load. 
The bias of our predictions in the first epoch was then calculated,
through comparison of the predictions with the actual path values
$y$, which follow from the reconstructed link values via $y=Gx$.
This correction was then used to adjust all other predictions
in the other 431 epochs, which amounts to simply shifting each 
curve in Figure~\ref{fig:GC-preds} upward a certain amount.
Boxplots of the relative bias remaining after application of this procedure
are shown in Figure~\ref{fig:bias-corrected-ts}. 
The predictions are now extremely accurate, being off usually by less than 
$0.3\%$, and almost always within $1\%$ --- even when
as few as only $k=3$ paths are measured.

\subsection{Anomaly Detection}
\label{sec:anomaly-detection}

The application in Section~\ref{sec:network-wide-average} evaluates
the predictor by standard statistical summaries, in essence looking at
the accuracy of the predictor at hitting an unknown target. But it is
also important to evaluate the accuracy in terms of accomplishing
higher-level tasks. One such higher-level task of importance is the
detection of potentially anomalous events.

\begin{figure}[tbp]
  \centering
  \includegraphics{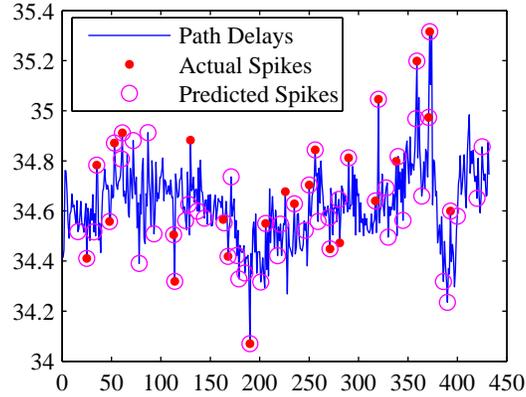}
  \caption{Comparison of predicted and actual spikes. The real spikes
  are those that exceed 3 times the standard deviation of the previous
  6 epochs. The predicted spikes are those epochs where the rank 9
  prediction exceeds 2 times the standard deviation of its previous 6
  epochs. }
  \label{fig:spikes}
\end{figure}

For the purposes of this application, we will define an anomaly as a
spike in the delay time that deviates from the average of the previous six
measurements (corresponding to one hour) 
by more than a specified threshold. For example, the red
dots in Figure~\ref{fig:spikes} indicate points at which the mean path
delay exceeds the mean of the previous six epochs by more than three
standard deviations. 

To predict when such anomalies occur, we look for spikes in the
predicted values. These predictions are highly correlated with the
actual mean path delays. For example, the $k=9$ prediction exhibits a
correlation coefficient of $\rho=0.930$. Recall that while a strong
correlation implies a strongly linear relation between the prediction
and the actual values, this does not mean the dynamic range
of the two signals are identical.   Thus we find it useful to explore
the effects of both $k$ and the detection threshold applied to the
predicted values.

\begin{figure}[tbp]
  \centering
  \includegraphics{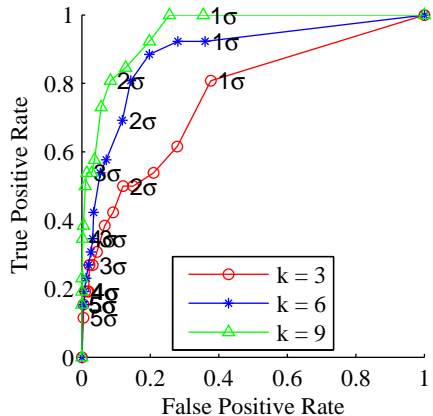}
  \caption{ROC curves for predicting $3\sigma$ spikes. The threshold
  used to predict the spikes is varied from $1\sigma$ 
    to $5\sigma$ in increments of $0.25\sigma$. }
  \label{fig:ROC-threshold}
\end{figure}

Insight on  the choice of
$k$ and threshold for the predictor can be obtained by examining
ROC (Receiver Operating Characteristic) curves such as those in
Figure~\ref{fig:ROC-threshold}. These plots of the true
positive rate against the false positive rate for different parameter
values are a common tool for establishing cutoff values for detection
tests.  Each curve in Figure~\ref{fig:ROC-threshold} is formed by
taking one value for $k$ and varying the threshold level.  Examining
these curves, one sees that for a given threshold, say $1\sigma$, the
true positive rate increases with the sample size $k$ while the false
positive rate stays about the same. Working with a $k=9$ prediction,
we see that the upper-left corner of the ROC curve (the best trade-off
between a low false positive rate and a high true positive rate)
occurs at around $2\sigma$. 

In Figure~\ref{fig:spikes}, the results are shown for the case 
$k=9$, with a threshold of $2\sigma$.
Circles have been placed along the actual path
delay time series at the epochs that were flagged as anomalies in the
predicted time series.  On the whole,
this predictor is quite accurate. Most of the major spikes are flagged,
resulting in a true positive rate of 81\%, while the false positive
rate is only 8\%.  Furthermore, most of these false positives seem to
occur at lesser spikes in the actual delay data.

\subsection{Subnetwork Comparison}
\label{sec:sub-netw-comp}
The applications in Sections~\ref{sec:network-wide-average}
and~\ref{sec:anomaly-detection} both involved predictions
of an average over the entire Abilene network.  However, it is
important to note that there is nothing in our framework that requires
that it be applied to a full network.  It applies equally to arbitrary
subnetworks and, as we will show now, can be particularly useful
for the task of subnetwork comparison.

Specifically, consider the problem of comparing the average delay times 
for the paths in
two groups, each consisting of all paths that originate at a
given node. 
This is an increasingly common scenario in the Internet. For example, many
customers now seek to optimize their access to the Internet via multihoming,
and choosing the best network ingress connection dynamically based on changing
network conditions. Another case occurs when providers peer at multiple
locations and thus have more than one ingress to a peer network. 

Let $\P_1$ and $\P_2$ denote the two sets of paths that we wish to
compare. 
If these measurements are being taken by a customer choosing
between two providers, then the only paths available for sampling are
probably those paths that belong to either $\P_1$ or $\P_2$.  To
ensure that we choose only from among those paths, we
restrict the routing matrix to a matrix $\tilde G$ which contains only
the rows of $G$ that correspond to paths in $\P_1$ and $\P_2$.

To determine which collection of paths has a shorter mean delay, we
define $l$ via
\begin{equation}
  \label{eq:15}
  l_i = 
  \begin{cases}
    1/\abs{\P_1} & \text{ if path $i$ is in } \P_1 \\
    -1/\abs{\P_2} & \text{ if path $i$ is in } \P_2 \text{,}
  \end{cases}
\end{equation}
so that $l^Ty$ is the difference between the mean delay of the paths
in $\P_1$ and the mean delay of the paths in $\P_2$.
The sign of $l^T y$ is therefore of interest. When $l^Ty$
is negative, $\P_1$ has a lower average delay, and when $l^T y$ is
positive, $\P_2$ has the lower average delay. To predict the sign of
$l^Ty$ we can use (\ref{eq:9}) to compute $\ahat^T y_s$ and check its
sign.

\begin{figure}[tbp]
  \centering
  \includegraphics{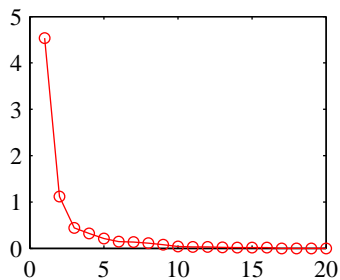}
  \caption{Spectrum of $\tilde GC$ for the subnetwork comparison
    between Chicago and Atlanta.}
  \label{fig:subnet-spec}
\end{figure}

To illustrate, we chose Chicago and Atlanta as our two ingress
points.
The spectrum of $\tilde GC$, shown in
Figure~\ref{fig:subnet-spec}, has a strong decay with a knee somewhere
around $k=5$. So our path selection method was applied to $\tilde GC$
with $k=5$, and we used the measurements on these five paths to predict
the difference in mean path delays out of Chicago and Atlanta.

\begin{figure}[tbp]
  \centering
  \includegraphics{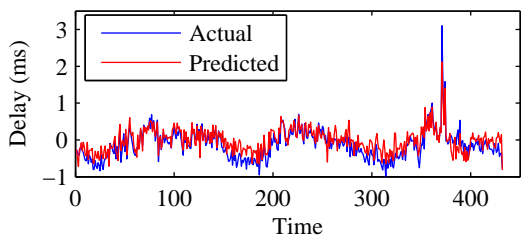}
  \caption{Subnetwork comparison between paths starting at Chicago and
    those starting at Atlanta. Prediction from a
    sample of 5 paths, with a correction for bias.}
  \label{fig:net-comp-pred}
\end{figure}

In Figure~\ref{fig:net-comp-pred}, we compare the predicted delay
difference against the actual delay difference.  The predictor tracks
the actual delay difference very well, with a correlation coefficient
of $\rho = 0.866$, and the ingress point with a shorter mean delay is
predicted (via the sign of $\ahat^T y_s$) correctly 79.6\% of the
time.

\begin{figure}[tbp]
  \centering
  \includegraphics{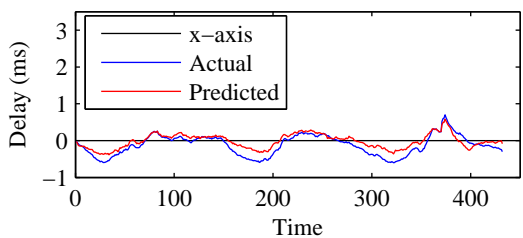}
  \caption{Exponential smoothing ($\alpha=0.1$) of the 
        predicted and actual time series in 
        Figure~\ref{fig:net-comp-pred}.}
  \label{fig:net-comp-pred-smooth}
\end{figure}

A good part of this roughly $11\%$ rate of error 
is due to the noise in the time-series.
A multihomed user is not likely to switch between providers on a
minute-to-minute basis, but rather might base decisions on 
predictions in a window of some length of time into the past.
To approximate this behavior, we apply an exponential smoothing
(with $\alpha=0.1$) to both the predicted and actual differences
in average delays.  As can
be seen in Figure~\ref{fig:net-comp-pred-smooth}, the prediction
tracks the actual delay difference very well, and the ingress point
with a shorter
mean delay is now correctly predicted over 88\% of the time.

We looked at the times at which the smoothed predictor and actual
delay differ in sign, and found that such periods fall into two
categories. First, there are minor periods where one series will dip
below or above zero for no more than an hour.  Alternatively,
the predictor will be slightly early or late in changing sign. The
offset is usually less than an hour, other than one notable exception
between $t=267$ and $t=283$ where the actual difference hovers just
below zero.

\begin{figure}[tbp] 
  \centering
  \includegraphics{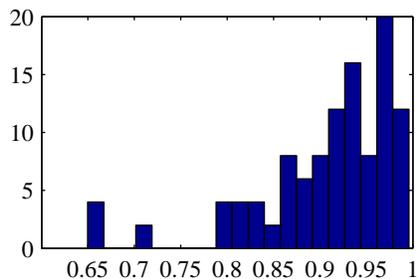}
  \caption{Correlations between unsmoothed predicted and actual
    differences in the mean delay for subnetworks. All predictions use
    $k=5$.} 
  \label{fig:subnet-corrs}
\end{figure}

Lastly, we point out that this performance in some sense carries
even more weight
when one considers the fact that this predictor for the 
Chicago/Atlanta 
pairing has a correlation with the actual difference
that is in the bottom quartile, over all pairs of ingress
points, as can be seen by examining Figure~\ref{fig:subnet-corrs}. 

\section{Conclusion}
\label{sec:conclusion}

In this paper we have demonstrated, using real data from an operating
network, that it is possible to monitor
end-to-end network delay properties quite accurately using a 
subset of paths substantially smaller than that needed for exact
monitoring.  We have illustrated the relevance of our approach
through the development of three practical applications of interest to 
network providers and customers: monitoring the `health' of a network,
anomaly detection, and comparison of subnetworks.

The framework proposed in this paper casts the problem of 
efficient network monitoring as one of statistical prediction.
Our approach exploits an observed redundancy
in links traversed by paths in real networks.  We explored 
the characteristics of this redundancy and found
them to be robust to 
variations in the state of the network,
such as link failure.

Our next steps include evaluation of our methods in a larger live network such
as PlanetLab. 
Additionally, it would be of interest to better understand the factors
underlying the observed path redundancy in the networks we have
examined, and the manner in which they contribute to the decay of
the eigenspectra of the corresponding routing matrices.  Initial
insight into such issues is beginning to emerge in works such 
as~\cite{chen04:algebraic}.

\section{Acknowledgements}
\label{sec:acknowledgements}
The authors would like to thank 
Anukool Lakhina (Computer Science Department, Boston University)
for his help with collecting and processing the data. 

\bibliographystyle{abbrv}
\bibliography{amp-sigmetrics}

\appendix
\section{Abilene Edges}
\label{sec:abilene-edges}

Table~\ref{tab:edge-nums} lists the edges of Abilene in the order that they
appear in Figures~\ref{fig:first-two-sing-vec-G-vs-GC} and
\ref{fig:delSpecEnergyDistrib}.

\begin{table}[htbp]
  \centering
  \begin{tabular}[b]{|rl|} \hline
\# & Edge \\ \hline
 1 & New York--Chicago \\
 2 & New York--Washington~D.C. \\
 3 & Chicago--New York \\
 4 & Chicago--Indianapolis \\
 5 & Washington~D.C.--New York \\
 6 & Washington~D.C.--Atlanta \\
 7 & Atlanta--Washington~D.C.\\
 8 & Atlanta--Indianapolis\\
 9 & Atlanta--Houston\\
10 & Indianapolis--Chicago\\
11 & Indianapolis--Atlanta\\
12 & Indianapolis--Kansas City\\
13 & Houston--Atlanta\\
14 & Houston--Kansas City\\
15 & Houston--Los Angeles\\
16 & Kansas City--Indianapolis \\
17 & Kansas City--Houston \\
18 & Kansas City--Denver \\
19 & Kansas City--Sunnyvale \\
20 & Denver--Kansas City \\
21 & Denver--Sunnyvale \\
22 & Denver--Seattle \\
23 & Sunnyvale--Kansas City \\
24 & Sunnyvale--Denver \\
25 & Sunnyvale--Los Angeles \\
26 & Sunnyvale--Seattle \\
27 & Los Angeles--Houston \\
28 & Los Angeles--Sunnyvale \\
29 & Seattle--Denver \\
30 & Seattle--Sunnyvale \\ \hline
\end{tabular}
 \caption{Abilene's directed edges.}
 \label{tab:edge-nums}
\end{table}

\end{document}